\documentclass[%
reprint,
superscriptaddress,
preprint,
 amsmath,amssymb,
 aps,
prb,
onecolumn
]{revtex4-2}

\usepackage{graphicx}
\usepackage{dcolumn}
\usepackage{bm}

\usepackage{xcolor}
\usepackage{multirow}
\usepackage{amsmath}
\usepackage{amssymb}
\usepackage{caption}
\usepackage{siunitx}
\usepackage{booktabs}
\begin{document}

\title{\textbf{Thermal Conductivity Limits of MoS$_2$ and MoSe$_2$: Revisiting High-Order Anharmonic Lattice Dynamics with Machine Learning Potentials}}

\author{Tu\u{g}bey Kocaba\c{s}}
\affiliation{Department of Advanced Technologies, Eskisehir Technical University, 26555 Eskisehir, T\"{u}rkiye}
\author{Murat Ke\c{c}eli}
\affiliation{Computational Science Division, Argonne National Laboratory, Lemont IL 60517, USA}
\author{Tanju Gürel}
\affiliation{Department of Physics, Tekirdağ Namik Kemal University, Tekirdağ TR 59030, T\"{u}rkiye}
\author{Milorad V. Milo\v{s}evi\'c}
\affiliation{COMMIT, Department of Physics, University of Antwerp, Groenenborgerlaan 171, B-2020 Antwerp, Belgium}
\author{Cem Sevik}
\affiliation{COMMIT, Department of Physics, University of Antwerp, Groenenborgerlaan 171, B-2020 Antwerp, Belgium}

\date{\today}

\begin{abstract}
Group-VI transition metal dichalcogenides (TMDs), MoS$_2$ and MoSe$_2$, have emerged as prototypical low-dimensional systems with distinctive phononic and electronic properties, making them attractive for applications in nanoelectronics, optoelectronics, and thermoelectrics. 
Yet, their reported lattice thermal conductivities ($\kappa$) remain highly inconsistent, with experimental values and theoretical predictions differing by more than an order of magnitude. 
These discrepancies stem from uncertainties in measurement techniques, variations in computational protocols, and ambiguities in the treatment of higher-order anharmonic processes. 
In this study, we critically review these inconsistencies, first by mapping the spread of experimental and modeling results, and then by identifying the methodological origins of divergence. 
To this end, we bridge first-principles calculations, molecular dynamics simulations, and state-of-the-art machine learning force fields (MLFFs) including recently developed foundation models. 
We train and benchmark GAP, MACE, NEP, and \textsc{HIPHIVE} against density functional theory (DFT) and rigorously evaluate the impact of third- and fourth-order phonon scattering processes on $\kappa$. 
The computational efficiency of MLFFs enables us to extend convergence tests beyond conventional limits and to validate predictions through homogeneous nonequilibrium molecular dynamics as well. 
Our analysis demonstrates that, contrary to some recent claims, fully converged four-phonon processes contribute negligibly to the intrinsic thermal conductivity of both MoS$_2$ and MoSe$_2$. 
These findings not only refine the intrinsic transport limits of 2D TMDs but also establish MLFF-based approaches as a robust and scalable framework for predictive modeling of phonon-mediated thermal transport in low-dimensional materials.

\end{abstract}

\maketitle

\section{Introduction}
Efficient thermal management has become a critical constraint in the design and operation of modern nanoelectronic, optoelectronic, and thermoelectric devices for energy conversion. 
As device dimensions continue to shrink and power densities increase, the ability of materials to dissipate heat effectively is essential to ensure optimal performance, stability, and long-term reliability. 
In this context, the lattice thermal conductivity ($\kappa$) has emerged as a key figure of merit for the selection and optimization of functional materials~\cite{Cahill2014JAP, Pop2012NR}.

Among emerging low-dimensional systems, two-dimensional (2D) materials, particularly group-VI transition metal dichalcogenides (TMDs) such as molybdenum disulfide (MoS$_2$) and molybdenum diselenide (MoSe$_2$), have attracted considerable attention due to their unique thermal, electronic, and optoelectronic properties~\cite{Yan2014, PhysRevLett.105.136805, Splendiani2010, Zhang2015, Picker2024, Bui2023, Park2024}. 
Significant experimental efforts have focused on quantifying their thermal conductivity using techniques such as Raman thermometry and optothermal methods~\cite{SaletaReig2022, Zhang2015}. 
However, even for the prototypical case of monolayer MoS$_2$, experimental values of $\kappa$ remain widely scattered. 
Reported room-temperature measurements range from $13~\text{W}\text{m}^{-1}\text{K}^{-1}$, based on early Raman studies on suspended flakes~\cite{Bae2017}, to $84~\text{W}\text{m}^{-1}\text{K}^{-1}$, using refined optical calibration techniques~\cite{Zhang2015}. 
Additional reports that account for anisotropic effects further extend the range from $24$ to $100~\text{W}\text{m}^{-1}\text{K}^{-1}$~\cite{Yang2020}. 
Similar variability is observed for MoSe$_2$~\cite{SaletaReig2022}, albeit with consistently lower values, often attributed to its heavier chalcogen atom and reduced phonon group velocities. 
The reported conductivities for MoSe$_2$ range from $20$~\cite{SaletaReig2022} to $59~\text{W}\text{m}^{-1}\text{K}^{-1}$~\cite{Zhang2015}.

In addition to experimental studies, first-principles calculations have been widely used to predict the thermal conductivity of these materials~\cite{RevModPhys.90.041002, nano13010117}. 
The combination of density functional theory (DFT) with the Peierls-Boltzmann transport equation (PBTE) has become the standard computational framework for such investigations. 
However, the resulting predictions exhibit substantial variation across the literature, largely due to methodological differences. 
Key contributing factors include the choice of exchange-correlation functional, supercell size, Brillouin zone sampling density, treatment of anharmonic force constants, and convergence criteria. 
For monolayer MoS$_2$, DFT-BTE-based estimates of $\kappa$ range from approximately $25$ to more than $150~\text{W}\text{m}^{-1}\text{K}^{-1}$, depending on the computational parameters employed. 
Similarly, the theoretical predictions for MoSe$_2$ vary between $17$ and $70~\text{W}\text{m}^{-1}\text{K}^{-1}$ (see Table \ref{tab:kappa_lit}).

Complementary molecular dynamics (MD) simulations have also been applied to study thermal transport in these materials~\cite{Ding2015, Hong2016, Jiang2025, Jin2015, Kandemir2016, Liu2013, Ma2020, Mobaraki2019, Wang2016, Xu2019, Zhang2017-2}.
However, the variability in the reported results is often as large as or greater than that seen in first-principles studies. 
For monolayer MoS$_2$, room-temperature values of $\kappa$ from MD simulations range from $1.35~\text{W}\text{m}^{-1}\text{K}^{-1}$~\cite{Liu2013} to $531\text{W}\text{m}^{-1}\text{K}^{-1}$~\cite{Xu2019}, while values for MoSe$_2$ span from $17.76$~\cite{Ma2020} to $76.2~\text{W}\text{m}^{-1}\text{K}^{-1}$~\cite{Jiang2025}. 
These large discrepancies are primarily attributed to differences in the employed interatomic potentials, but other factors, including system size, boundary conditions, thermostat algorithms (e.g., Langevin or Nosé–Hoover), and averaging times, also play significant roles. 

To address these discrepancies, increasing attention has been given to the role of higher-order phonon scattering processes, specifically those beyond the standard three-phonon interactions typically included in DFT-based thermal transport calculations~\cite{Kocaba2024, Chaudhuri2024, Kagdada2025, Sun2023, Batista2025, Feng2017, Zhang2022, Li2022, PhysRevB.97.045202, PhysRevB.100.245203}.
Recent studies have shown that four-phonon processes can significantly reduce predicted thermal conductivities~\cite{Chaudhuri2024, Sun2023, Kagdada2025, Batista2025, Feng2017, Zhang2022, Li2022, PhysRevB.97.045202, PhysRevB.100.245203}, indicating that models limited to three-phonon interactions may systematically overestimate $\kappa$, particularly at elevated temperatures or in materials with strong anharmonicity. 
However, these calculations remain computationally challenging, as they demand large supercells and dense Brillouin zone sampling; consequently, the convergence and accuracy of reported fourth-order interactions are often uncertain.

In this work, we employ machine learning force fields (MLFFs) in conjunction with first-principles calculations and molecular dynamics simulations to advance the understanding of the fundamental limits in thermal transport of MoS$_2$ and MoSe$_2$ monolayers and address the aforementioned discrepancies. 
We systematically benchmark thermal conductivity predictions against density functional theory (DFT) by analyzing convergence with respect to both third- and fourth-order phonon scattering processes. 
Specifically, we assess the accuracy of four Multi-Layer Perceptron (MLP) frameworks (GAP, MACE, NEP, and \textsc{HIPHIVE}) in modeling the thermal transport of these materials. 
Upon establishing strong agreement with DFT results, we leverage the computational efficiency of MLFFs to explore effects beyond conventional reach, including higher-order neighbor interactions and four-phonon scattering. 
Additionally, we validate thermal conductivity estimates via homogeneous nonequilibrium molecular dynamics (HNEMD) simulations. 
These high-accuracy simulations, unattainable with standard DFT methods, highlight the critical role of machine learning in scalable modeling of complex phonon processes in low-dimensional materials. 
Importantly, our results demonstrate that fully converged fourth-order scattering contributions are negligible, in contrast to some earlier reports, and thereby refine the intrinsic thermal transport limits of monolayer transition metal dichalcogenides.

\section{Computational Details}

Structural relaxations and force calculations were performed using DFT within the Perdew–Burke–Ernzerhof (PBE) formulation of the generalized gradient approximation (GGA)~\cite{Perdew1996} as implemented in \textsc{VASP}~\cite{Kresse1993}. 
A plane-wave basis set with an energy cutoff of 600~eV was used for all the calculations. 
Brillouin zone sampling was performed using a $\Gamma$-centered Monkhorst--Pack~\cite{Monkhorst1976} mesh of $24 \times 24 \times 1$ $k$-points. 
The electronic self-consistency loop was converged to an energy difference of $10^{-6}$~eV, while ionic relaxations were terminated when forces fell below $10^{-2}$~eV/\AA. 
To eliminate spurious interactions between periodic images, a vacuum spacing of at least 20~\AA\ was applied along the out-of-plane ($z$) direction.

Training, validation, and test datasets for the MLFFs were generated using the \textit{on-the-fly} learning scheme in \textsc{VASP}~\cite{Jinnouchi2019}. 
To capture all relevant phonon interactions, molecular dynamics simulations were carried out at multiple temperatures. 
The same computational parameters as those used in the first-principles calculations were applied to maintain consistency in data quality. 
For each of MoS$_2$ and MoSe$_2$, 3000 configurations were selected for the training dataset, while 250 structures were used for both the validation and test datasets.

Interatomic force constants (IFCs) up to fourth order were evaluated to accurately capture anharmonic lattice dynamics. 
The second- and third-order IFCs from DFT were obtained via the finite displacement approach using \textsc{VASP} in combination with the \textsc{Phonopy}~\cite{Togo2015} and \texttt{thirdorder.py}~\cite{Li2012} packages, respectively. 
For this purpose, an $8 \times 8 \times 1$ supercell and a $4 \times 4 \times 1$ $\Gamma$-centered \textit{k}-point mesh were adopted. 
To ensure high numerical accuracy in the force constants, the electronic minimization tolerance was tightened to $10^{-8}$~eV. 
The interaction range was extended up to the 18$^\mathrm{th}$ nearest neighbors (NNs) for third-order IFCs.

Fourth-order IFCs were obtained exclusively from MLFFs since DFT calculations at this level are computationally prohibitive.
For comparison, third-order IFC calculations with an 18-NN cutoff required 1,144 force evaluations, whereas fourth-order IFC calculations with a 10-NN cutoff demanded nearly 35{,}000.
This stark increase in computational cost illustrates the impracticality of DFT-based fourth-order calculations, which justifies the use of MLFFs for this purpose. 
Convergence of fourth-order displacements was examined for amplitudes between 0.01 to 0.05~\AA, with 0.04~\AA\ selected (see Supplementary Material). 
Similarly, interaction ranges were tested up to the 10$^{\text{th}}$ NN and convergence was achieved at the 6$^{\text{th}}$ NN, which was used in all subsequent calculations.

The MLFFs employed in this study include Gaussian Approximation Potentials (GAP)~\cite{Bartok2010, gabor2015}, MACE~\cite{Batatia2025, Batatia2022Design}, NeuroEvolution Potential (NEP)~\cite{Song2024}, and \textsc{HIPHIVE}~\cite{Eriksson2019}.  
For \textsc{HIPHIVE}, 200 DFT-generated training configurations were used for each of monolayer MoS$_2$ and MoSe$_2$, employing the same computational parameters described above to ensure data consistency.  
During the fitting procedure, interaction cutoffs were set individually for $n$-body terms up to six-body interactions. For MoS$_2$, the cutoffs for 2-, 3-, 4-, 5-, and 6-body interactions were 12.5, 12.5, 6.0, 4.5, and 3.0~\AA, respectively, whereas for MoSe$_2$, the corresponding values were 13.28, 13.28, 6.5, 5.0, and 3.0~\AA. These cutoffs were selected based on convergence tests of the predicted force constants and phonon dispersion curves.  
The fitting parameters used for GAP, MACE, and NEP are provided in the Supplementary Material, and the resulting potential files have been deposited in a public repository (Zenodo, DOI: \texttt{XXXX}).  

To ensure the reliability of higher-order IFC predictions, each MLFF was first validated by comparing its second-order IFC-derived phonon dispersion curves and third-order IFC-derived lattice thermal conductivity values with DFT results, which were used as reference. Once satisfactory agreement with the DFT benchmarks was achieved, the validated MLFFs were used to compute the fourth-order IFCs. 
This strategy ensured that higher-order anharmonic effects were included in a physically consistent and computationally efficient manner.

Lattice thermal conductivity and related transport properties, including phonon lifetimes and Grüneisen parameters, were computed by iteratively solving the Peierls–Boltzmann transport equation (PBTE)~\cite{Ziman1960} using the standalone \textsc{FourPhonon} package~\cite{HAN2022108179}, an extension of \textsc{ShengBTE}~\cite{Li2014} that supports both three- and four-phonon scattering processes. 
A dense $q$-mesh of $80 \times 80 \times 1$ was employed for convergence, and up to 18$^\mathrm{th}$ NN interactions were included in the three-phonon scattering calculations. For four-phonon scattering, the same $q$-grid was used, and the scattering rates were evaluated using the maximum likelihood estimation (MLE) method implemented in \textsc{FourPhonon}.
To ensure statistical accuracy, 100{,}000 sampling points were used for both the phase space estimation and the scattering rate integration. 
The effective layer thicknesses were taken as 6.15~\AA\ and 6.47~\AA\ for monolayer MoS$_2$ and MoSe$_2$, respectively.

HNEMD simulations were performed using the \textsc{GPUMD}~\cite{Xu2025, Fan2019, Gabourie2021} package with NEP to evaluate the lattice thermal conductivity at the molecular dynamics level. 
For both materials, square simulation cells were constructed, each containing 37{,}440 atoms. The lattice constants were set to 330~\AA\ for MoS$_2$ and 344~\AA\ for MoSe$_2$. Periodic boundary conditions were applied in the plane, and all simulations used a time step of 1~fs. 
At each target temperature (300~K and 600~K), the systems were equilibrated for 1{,}000{,}000 steps (1~ns) in the NVT ensemble using a Nosé–Hoover chain thermostat. 
Following equilibration, a small external driving force was applied to generate a homogeneous heat current, and the system was propagated for an additional 10{,}000{,}000 steps (10~ns) in the nonequilibrium production stage. 
An effective thickness of 6.15~\AA\ for MoS$_2$ and 6.47~\AA\ for MoSe$_2$ was used for volume normalization in the thermal conductivity calculations. 
For each material and each temperature, the thermal conductivity was determined by averaging the results of 50 independent simulations.

\section{Results and Discussion}
Table~\ref{tab:kappa_lit} summarizes reported values of room-temperature lattice thermal conductivity for monolayer MoS$_2$ and MoSe$_2$, highlighting the significant discrepancies present in the literature. 
To allow a meaningful comparison with our results, we have rescaled the reported $\kappa$ values using consistent out-of-plane lattice constant values, as specified in the Computational Details section.
This scaling is essential because the out-of-plane lattice constant for two-dimensional materials is arbitrary and determines the final value of thermal conductivity in units of Wm$^{-1}$K$^{-1}$. 
Here, we particularly note that for entries presented in italic font, the original out-of-plane lattice constant values were not reported in the corresponding publication. 

As shown in Table~\ref{tab:kappa_lit}, inconsistencies appear not only in first-principles calculations but also in molecular dynamics simulations and experimental measurements. 
For both materials, a reliable comparison with the experimental data remains challenging. 
While some variability in experimental values is expected due to difficulties in fabricating high-quality monolayers and accurately measuring thermal transport at the nanoscale, the divergence observed within results from the same theoretical approach is often attributable to differences in computational implementation. 
In the case of MD simulations, the results strongly depend on the quality of the employed interatomic potentials. 
Classical potentials, which are commonly used, often lack the fidelity required to reproduce first-principles results.

We leverage recent advances in machine learning force fields to overcome the limitations of both first-principles and classical MD approaches. 
Our results demonstrate that MLFFs enable high-accuracy modeling of lattice thermal transport in two-dimensional materials and allow for an efficient exploration of the intrinsic limits of thermal conductivity. 
The following sections present the framework we applied for this purpose and the corresponding results in detail.

\begin{table}[h!]
\scriptsize
\renewcommand{\arraystretch}{0.85} 
\setlength{\tabcolsep}{8pt} 
\caption{Room-temperature lattice thermal conductivity values of monolayer MoS$_2$ and MoSe$_2$ collected from the literature, based on first-principles, molecular dynamics, and experimental studies. All reported values have been rescaled according to the out-of-plane lattice constants used in this work: 6.15~\AA{} for MoS$_2$ and 6.47~\AA{} for MoSe$_2$. Italicized values indicate that the original source did not provide an explicit out-of-plane lattice constant.}
\centering
\begin{tabular}{lclc}
\hline
\hline
\multicolumn{4}{c}{\textit{First-Principles}}\\
\multicolumn{2}{c}{\textbf{MoS$_2$}} & \multicolumn{2}{c}{\textbf{MoSe$_2$}}\\
\hline
\textbf{Method} & $\mathbf{\kappa}$ ($\text{Wm}^{-1}\text{K}^{-1}$) & \textbf{Method} & $\mathbf{\kappa}$ ($\text{Wm}^{-1}\text{K}^{-1}$)  \\
\hline
DFT-BTE~\cite{Peng2016} & 151.36 & DFT-BTE~\cite{Farris2024} & 54.13 \\
DFT-BTE~\cite{Gu2016} & 135.20 & DFT-BTE~\cite{Kumar2015} & \textit{$\sim$70} \\
DFT-BTE~\cite{Gandi2016} & 130.20 & DFT-BTE~\cite{Zulfiqar2019} & \textit{$\sim$60} \\
DFT-BTE~\cite{Zhao2018} & 130.00 & DFT-BTE~\cite{Gu2014} & \textit{54} \\
DFT-BTE~\cite{Gu2014} & 103.00 & DFT-BTE~\cite{Zhang2017} & \textit{46.2} \\
DFT-BTE~\cite{Li2013} & 81.42 & DFT-DFPT-Slack Model~\cite{Peng2016-2} & \textit{17.6} \\
DFT-BTE~\cite{Farris2024} & 89.56 \\
DFT-DFPT-NEGF~\cite{Cai2014} & 24.52 \\
DFT-BTE ($3^{\mathrm{ph}}$, $3^{\mathrm{ph}}$+$4^{\mathrm{ph}}$)~\cite{xhprb} & \textit{133.5, 27.7} \\
DFT-BTE~\cite{Zhang2017} & \textit{82.2} \\
DFT-BTE~\cite{Zulfiqar2019} & \textit{$\sim$75} \\
DFT-DFPT-Slack Model~\cite{Peng2016-2} & \textit{33.6} \\
DFT-DFPT-Umklapp Model~\cite{Su2015} & \textit{29.2} \\
\hline
\hline
\multicolumn{4}{c}{\textit{Molecular Dynamics}} \\
\hline
REBO-LJ-HNEMD~\cite{Xu2019} & 123.66 & SW-NEMD~\cite{Zhang2017-2} & 24.80\\
(SW13, SW13E, SW16)-HNEMD~\cite{Xu2019} & 535.85, 203.98, 290.65 & SW-EMD-Green-Kubo~\cite{Kandemir2016} & 40.19\\
TB-(EMD-NEMD)~\cite{Liu2013} & 0.97, 1.22 & SW-NEMD-(AC, ZZ)~\cite{Ma2020} & 17.76, 18.93\\
SW-NEMD~\cite{Zhang2017-2} & 32.89 & SW-NEMD-(AC, ZZ)~\cite{Hong2016} & 43.88, 41.63\\
SW-RNEMD-(AC, ZZ)~\cite{Wang2016} & 32.95, 53.91 & MLFF-NEP-HNEMD~\cite{Jiang2025} & 77.73\\
SW-EMD-Green-Kubo~\cite{Kandemir2016} & 90.00 & SW-SED~\cite{Mobaraki2019} & \textit{29.18}\\
SW-EMD-Green-Kubo~\cite{Jin2015} & 116.99\\
SW-NEMD-(AC, ZZ)~\cite{Hong2016} & 101.39, 110.26\\
SW-NEMD~\cite{Ding2015} & 19.95\\
MLFF-NEP-HNEMD~\cite{Jiang2025} & 161.62\\
SW-SED~\cite{Mobaraki2019} & \textit{89.4}\\
\hline
\hline
\multicolumn{4}{c}{\textit{Experimental}}\\
\hline
Raman(Heat Diff. Modeling)~\cite{Yan2014} & 36.46 & Mech.Exf.-Raman(vacuum, air)~\cite{SaletaReig2022} & (\textit{20, 250}) \\
Raman(Heat Diff. Modeling)~\cite{Taube2015} & 70.80 & Mech.Exf.-Raman~\cite{Zhang2015} & \textit{59} \\
CVD-RTD~\cite{Yarali2017} & 30 \\
Mech.Exf.-Raman~\cite{Zhang2015} & \textit{84} \\
CVD-Opt. Mod.~\cite{Dolleman2018} & \textit{19.8} \\
CVD-LHD~\cite{Gu2016} & \textit{13.3} \\
CVD-MJH~\cite{Yang2020} & \textit{24-100} \\
\hline
\end{tabular}
\label{tab:kappa_lit}
\end{table}

\subsection{ML potentials}
\begin{figure*}[h!]
    \centering
    \includegraphics[width=1\textwidth]{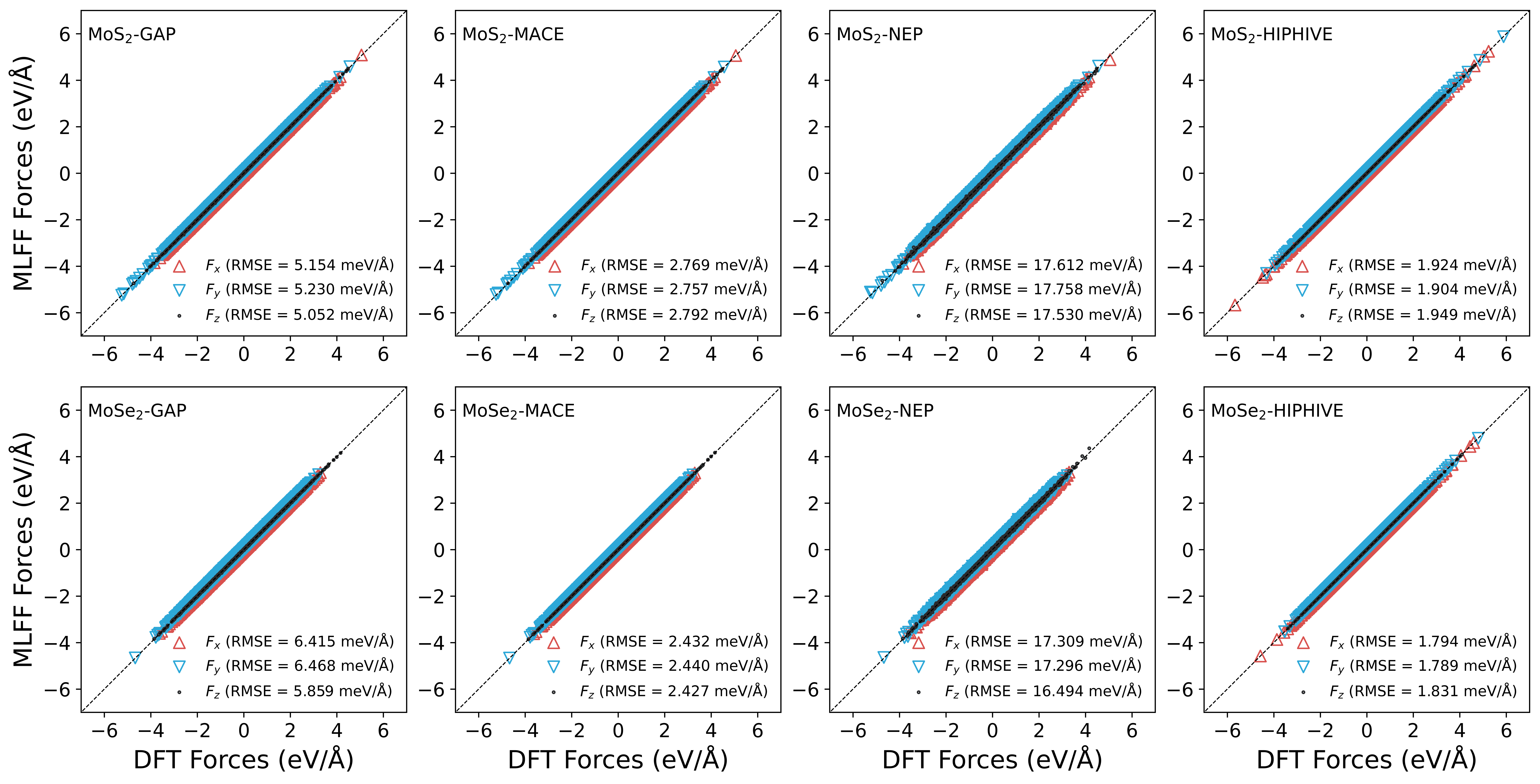}
    \caption{Comparison of MLFF-predicted atomic forces with DFT reference values for monolayer MoS$_2$ and MoSe$_2$ on an independent test set.}
    \label{fig:rmse_test}
\end{figure*}

 As a first step, we evaluated the accuracy of state-of-the-art machine learning force fields by comparing their predicted atomic forces with DFT-calculated reference values on an independent test data set. 
 The atomic forces obtained for each model (GAP, MACE, NEP, and \textsc{HIPHIVE}) are plotted against the DFT forces for monolayer MoS$_2$ and MoSe$_2$ in Figure~\ref{fig:rmse_test}. 
 Among these models, \textsc{HIPHIVE} achieves the lowest root-mean-square errors (RMSEs), with values consistently below $2~\mathrm{meV}/\mathrm{\AA}$ in all Cartesian directions. 
 This outcome is consistent with \textsc{HIPHIVE}’s specialized design as a Python library dedicated to efficiently extracting high-order anharmonic force constants from first-principles data. 
 However, we would like to note that while the data set structure differs for \textsc{HIPHIVE}, it was generated using the same settings and the precision level with other models, enabling a fair comparison. 
 GAP, MACE, and NEP were trained on the same DFT data set, allowing a direct comparison of their generalization capabilities. 
 GAP is based on a kernel regression framework that utilizes Smooth Overlap of Atomic Positions (SOAP) descriptors to encode local atomic environments. 
 MACE employs an equivariant message-passing neural network architecture designed to capture the geometric and symmetry properties of atomistic systems. 
 NEP, which can be used within the GPUMD package, uses symmetry-preserving neural networks optimized for GPU-accelerated simulations. 
 Despite differences in their internal architectures, all three models yield consistent results compared to DFT reference forces. 
 MACE and GAP produce lower RMSEs in the range of $2$-$6~\mathrm{meV}/\mathrm{\AA}$, while NEP shows higher values, around $17~\mathrm{meV}/\mathrm{\AA}$. 
 However, NEP delivers physically meaningful results, as further demonstrated by its accurate reproduction of both the phonon dispersions and the lattice thermal conductivity over a broad temperature range (details are presented below). 
 These findings demonstrate that NEP captures the key characteristics of interatomic interactions relevant to thermal transport.

\subsection{Thermal transport properties: PBTE solution via DFT and MLFFs}
\begin{figure*}[h!]
    \centering
    \includegraphics[width=0.9\textwidth]{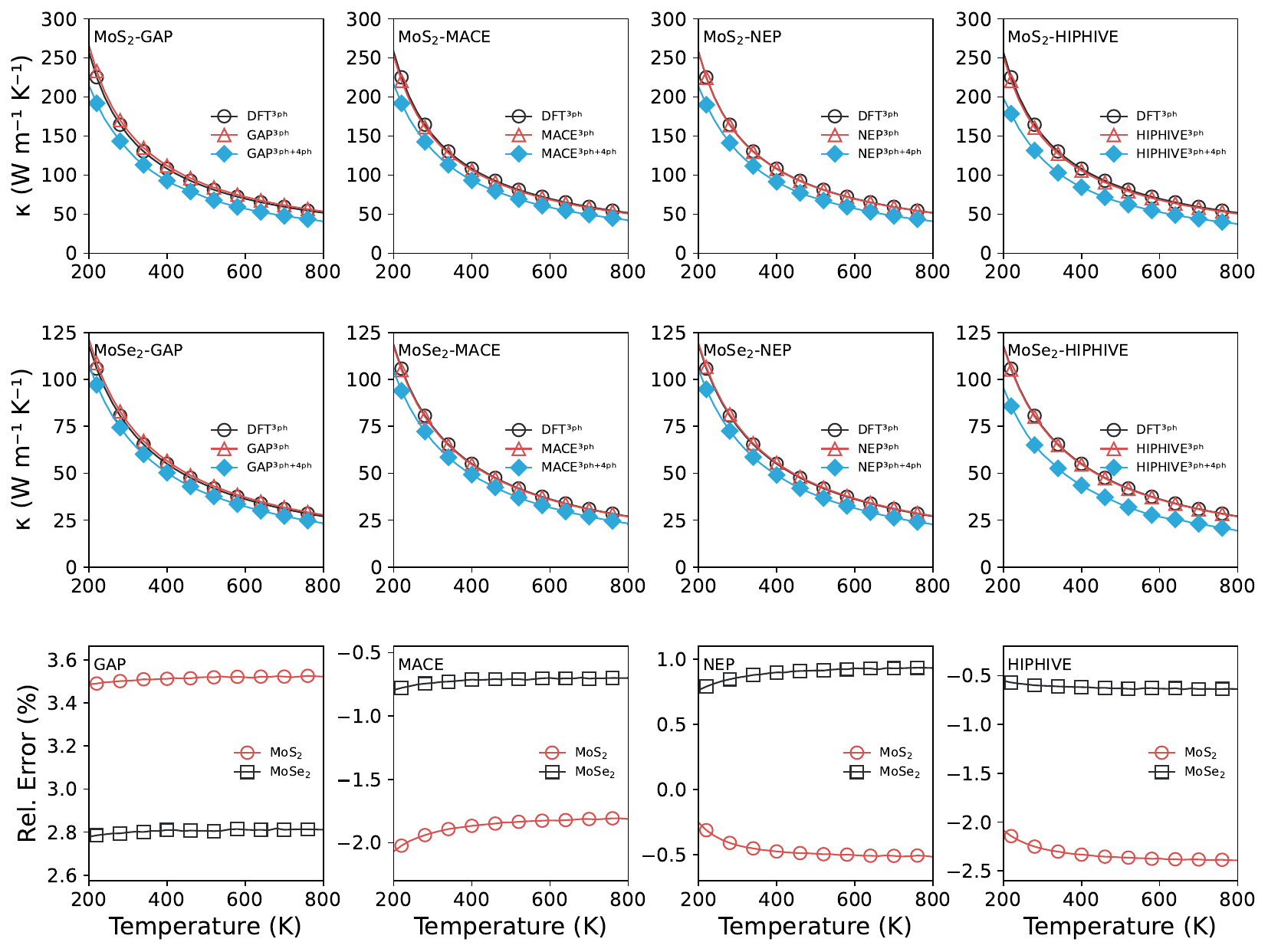}
    \caption{{Lattice thermal conductivity of MoS$_2$ and MoSe$_2$ as a function of temperature (200–800 K), calculated considering (i) three-phonon interactions up to the 13$^{\text{th}}$ nearest neighbors, and (ii) combined three-phonon (13$^{\text{th}}$ nearest neighbors) and four-phonon (6$^{\text{th}}$ nearest neighbors) interactions. The bottom panels show the relative errors of MLFF-predicted three-phonon thermal conductivities with respect to DFT values for both materials.}}
    \label{fig:tc}
\end{figure*}

The predictive performance of the GAP, MACE, NEP, and \textsc{HIPHIVE} potentials was evaluated by comparing their calculated lattice thermal conductivity values with DFT results for monolayer MoS$_2$ and MoSe$_2$ over the temperature range of 200--800~K. 
The comparison is illustrated in Figure~\ref{fig:tc}, which includes both the calculated conductivity values and the relative errors with respect to DFT. 
It is important to note that the thermal conductivity values were not predicted directly by the potentials themselves. 
Instead, atomic forces obtained from each potential were used to construct second- and third-order force constants, which were then employed in the iterative solution of the Peierls-Boltzmann transport equation. 

In all comparative $\kappa$ calculations presented in Figure~\ref{fig:tc}, phonon interactions were considered up to the 13$^{\mathrm{th}}$ NN shell, and the atomic displacement magnitude in the finite displacement method was set to 0.03~\AA\ for both second- and third-order force constant evaluations. 
All computations were performed using a $8 \times 8 \times 1$ supercell containing 192 atoms.
For the combined three-phonon and four-phonon (3$^{\mathrm{ph}}$+4$^{\mathrm{ph}}$) scattering processes, convergence tests—performed as strongly recommended in \cite{Zhou2023} — resulted in the use of a 0.04~\AA\ displacement and a 6$^{th}$ NN cutoff. 
In this comparison, three-phonon scattering processes were considered consistently across all models and the DFT reference. 

For MoS$_2$, NEP yields the closest agreement with DFT, with relative errors ($100[\kappa_{MLFF}-\kappa_{DFT}]\kappa_{DFT}^{-1}$ remaining below 0.6\% across the entire temperature range as seen in the bottom row of the Figure~\ref{fig:tc}. GAP shows the largest deviations, with errors up to 3.5\%, followed by HIPHIVE with deviations reaching 2.4\%. MACE also performs well, though its deviations can rise to about 2.1\%. For MoSe$_2$, the overall trend is similar: NEP and HIPHIVE both remain within 1\% relative error, while MACE shows modest deviations under 0.8\%. GAP again presents the largest differences, up to about 2.8\%. 
These comparisons indicate that all models produce reliable predictions of thermal conductivity, with NEP consistently showing the best agreement with DFT. The somewhat larger deviations for GAP, and to a lesser extent HIPHIVE and MACE, highlight subtle differences in force-matching accuracy. Such variations underscore the importance of careful potential selection when quantitative precision in monolayer thermal transport is required.

\begin{figure}[h!]
    \centering
    \includegraphics[width=0.5\columnwidth]{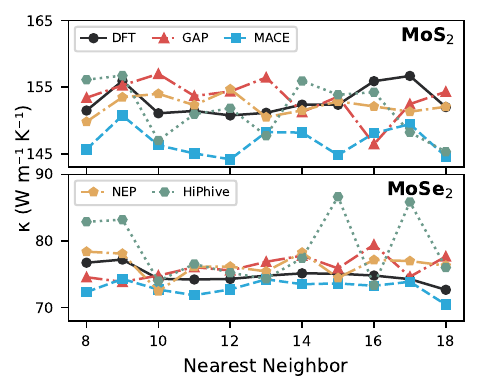}
    \caption{Lattice thermal conductivity ($\kappa$) of monolayer MoS$_2$ and MoSe$_2$ at room temperature as a function of the nearest-neighbor cutoff used in the third-order force constant calculations. Results are shown for DFT and machine-learned force fields (GAP, MACE, NEP, and \textsc{HIPHIVE}) including only three-phonon (3$^{\mathrm{ph}}$) scattering contributions.}
    \label{fig:nn_tc}
\end{figure}

For the 3$^{\mathrm{rd}}$-order calculations, the 13$^{\text{th}}$ NN cutoff was selected based on a systematic convergence analysis of  $\kappa$ as a function of the NN range; see Figure~\ref{fig:nn_tc}. 
Subsequently, this value was adopted as a reference point for the previous comparison of $\kappa$ in different interatomic potentials. 
However, the level of fluctuations beyond the 13$^{\text{th}}$ NN is not the same for different approaches, as illustrated in Figure~\ref{fig:nn_tc}. 
The more pronounced fluctuations observed in MLFFs predictions arise from both physical and numerical factors. 
Physically, $\kappa$ depends on the IFCs, and truncating interactions at different distances alters the included phonon-phonon scattering pathways. 
Numerically, MLFFs are data-driven models that interpolate atomic forces based on training data distributions. 
As the interaction cutoff increases, models must extrapolate to less-represented atomic environments, which can introduce instability or increased error, particularly when long-range interactions are weak but non-negligible. 
In contrast, DFT provides forces derived from a self-consistent quantum-mechanical formalism, which is inherently less sensitive to such extrapolation and better captures the behavior of interaction decay. 
The force prediction RMSEs presented in Figure~\ref{fig:rmse_test} and the further analysis presented below confirm that the MLFFs remain accurate within the trained range. 
The moderate fluctuation in $\kappa$ is not due to overfitting or extrapolation errors, as evidenced by consistent performance of the test set and strong agreement with DFT throughout the entire temperature range.

To ensure consistent and accurate thermal conductivity predictions using the solution of PBTE, we find it more appropriate to average the values over a small range extending slightly beyond the DFT convergence threshold. 
Specifically, the values of $\kappa$ at 300 and 600 K averaged over the 13$^{\text{th}}$–18$^{\text{th}}$ NNs range are summarized in Table~\ref{tab:tc_table_compact_mlff} (columns labeled 3$^\mathrm{ph}$), together with the relative errors referenced to DFT. 
This procedure yields room-temperature $\kappa$ values of 153 Wm$^{-1}$K$^{-1}$ for monolayer MoS$_2$ and 74 Wm$^{-1}$K$^{-1}$ for monolayer MoSe$_2$. 
Relative errors of less than 5\% for GAP, MACE, and NEP demonstrate that trained MLFFs achieve first-principles-level accuracy in predicting lattice thermal conductivity.

\begin{table}[htbp]
\centering
\caption{Lattice thermal conductivity values (Wm$^{-1}$K$^{-1}$) of monolayer MoS$_2$ and MoSe$_2$ at 300 and 600 K, calculated using the optimized MLFFs. Results labeled 3$^{\mathrm{ph}}$ are obtained from PBTE including only three-phonon scattering, with $\kappa$ values averaged over the 13$^{\text{th}}$–18$^{\text{th}}$ nearest-neighbor (NN) range to ensure convergence with DFT. The 4$^{\mathrm{ph}}$ results include both three- and four-phonon processes, where three-phonon contributions are obtained for the 13$^{\text{th}}$ NN and four-phonon contributions are averaged over the 6$^{\text{th}}$, 8$^{\text{th}}$, and 10$^{\text{th}}$ NN ranges.}
\label{tab:tc_table_compact_mlff}
\scriptsize
\setlength{\tabcolsep}{5pt}
\begin{tabular}{|l|l|r|r|r|r|}
\hline
\multirow{2}{*}{Material} & \multirow{2}{*}{Method}
& \multicolumn{2}{c|}{300~K} & \multicolumn{2}{c|}{600~K} \\
\cline{3-6}
& & 3$^{\mathrm{ph}}$& 4$^{\mathrm{ph}}$& 3$^{\mathrm{ph}}$& 4$^{\mathrm{ph}}$\\
\hline
\multirow{6}{*}{MoS\textsubscript{2}}
& DFT (ref)        & 153 (ref)     & \textemdash   & 71 (ref)   & \textemdash \\
& GAP              & 152 [$-0.65\%$] & 140 & 70 [$-1.41\%$] & 61 \\
& MACE             & 147 [$-3.92\%$] & 139 & 68 [$-4.23\%$] & 60 \\
& NEP              & 152 [$-0.65\%$] & 138 & 70 [$-1.41\%$] & 59 \\
& \textsc{HIPHIVE} & 151 [$-1.31\%$] & 127 & 70 [$-1.41\%$] & 53 \\
\hline
\multirow{6}{*}{MoSe\textsubscript{2}}
& DFT (ref)        & 74 (ref)      & \textemdash   & 36 (ref)   & \textemdash \\
& GAP              & 77 [$+4.05\%$] & 72   & 37 [$+2.78\%$] & 33 \\
& MACE             & 73 [$-1.35\%$] & 70 & 35 [$-2.78\%$] & 33 \\
& NEP              & 76 [$+2.70\%$] & 70 & 37 [$+2.78\%$] & 32 \\
& \textsc{HIPHIVE} & 79 [$+6.76\%$] & 63 & 38 [$+5.56\%$] & 28 \\
\hline
\end{tabular}
\normalsize
\end{table}

These well-converged and carefully benchmarked results that lie in the upper bound of the values reported in the first-principles literature: 81–151~Wm$^{-1}$K$^{-1}$ for MoS$_2$ and~18–75 Wm$^{-1}$K$^{-1}$ for MoSe$_2$, determine the reference point that can be obtained by using a similar first-principles framework.  
Notably, the calculated $\kappa$ values for both materials exceed most experimental reports, as summarized in Table~\ref{tab:kappa_lit}. 
The generally lower experimental values arise from several factors, including the inherent challenges of accurately measuring thermal conductivity in two-dimensional systems and, more critically, the unavoidable presence of structural imperfections. 
Even low concentrations of point defects, grain boundaries, or interactions with supporting substrates can introduce substantial phonon scattering, leading to significant reductions in the observed thermal conductivity~\cite{HUANG2023100210, PhysRevMaterials.2.095001, C9TA11424F, Ren2016}.

On the other hand, first-principles predictions that consider only third-order phonon-phonon scattering processes may significantly overestimate the lattice thermal conductivity. 
Several studies have shown that the inclusion of four-phonon scattering can lead to substantial reductions in the predicted $\kappa$ values. 
Notable examples include decreases from 3383 to 810 Wm$^{-1}$K$^{-1}$ in graphene~\cite{PhysRevB.97.045202}, 1303 to 180 Wm$^{-1}$K$^{-1}$ in monolayer $h$-BN~\cite{Batista2025}, 3322 to 1721 Wm$^{-1}$K$^{-1}$ in BAs~\cite{Yang2019}, 181 to 51 Wm$^{-1}$K$^{-1}$ in AlSb~\cite{Yang2019}, 421 to 210 Wm$^{-1}$K$^{-1}$ in BS~\cite{Kagdada2025}, 332 to 57 Wm$^{-1}$K$^{-1}$ in BSe~\cite{Kagdada2025}, and 109.25 to 11.67 Wm$^{-1}$K$^{-1}$ in Penta-NiN$_2$~\cite{Zhang2022}. 
For monolayer MoS$_2$, the material of interest in this work, Chaudhuri \textit{et al.} reported a reduction from 133.5 to 27.7 Wm$^{-1}$K$^{-1}$ upon accounting for four-phonon interactions~\cite{xhprb}. 

Motivated by these findings, we systematically examine the impact of fourth-order phonon-phonon scattering processes on the lattice thermal conductivity of the studied materials. Our approach combines an iterative solution of the Peierls-Boltzmann transport equation (PBTE) for three-phonon interactions with the relaxation time approximation (RTA) for four-phonon scattering as implemented in \textsc{FourPhonon} code. 
To enable accurate and computationally efficient evaluation of higher-order force constants, we employ MLFFs, which allow us to include extended interaction ranges—up to the 10$^{\mathrm{th}}$ nearest neighbor—encompassing approximately 35,000 distinct fourth-order atomic displacements. 
In contrast, conventional first-principles methods are severely constrained by computational cost, often limiting the interaction range and potentially leading to unconverged or inconsistent predictions of $\kappa$. 
As in the three-phonon case, the final conductivity values were averaged over the 6$^{\mathrm{th}}$, 8$^{\mathrm{th}}$, and 10$^{\mathrm{th}}$ NN cutoffs to ensure convergence and robustness (see Figure~S8 in the Supporting Information).

As shown in Figure~\ref{fig:tc}, inclusion of four-phonon scattering yields a moderate reduction of smaller than 20\% in the lattice thermal conductivity. 
At room temperature, the final values obtained after accounting for fourth-order phonon interactions are about 140 and 70 W m$^{-1}$ K$^{-1}$ for MoS\textsubscript{2} and MoSe\textsubscript{2}, respectively. 
The corresponding reductions at 300 and 600 K are summarized in Table\ref{tab:tc_table_compact_mlff} (columns labeled 4$^{\mathrm{ph}}$) for all MLFFs considered. 
Although this reduction is consistently observed across models, it is significantly less pronounced than the drastic suppression reported in earlier DFT-based studies (e.g., from 133.5 to 27.7 W m$^{-1}$ K$^{-1}$ in Ref.~\cite{xhprb}). 
These results highlight the importance of employing highly accurate MLFFs to obtain converged higher-order phonon scattering rates, thereby ensuring reliable and physically meaningful predictions of lattice thermal conductivity.

To elucidate the role of anharmonic phonon scattering further, we systematically analysed phonon lifetimes in monolayer MoS$_2$ and MoSe$_2$ at T=300 K arising from both three-phonon and four-phonon interactions. 
These lifetimes were computed via forces obtained from both DFT and MLFFs. 
As shown in Figures~S1-S4, 3$^{\mathrm{ph}}$ lifetimes in MoS$_2$ reach up to 250~ps in the low-frequency acoustic regime ($<$3~THz) and decrease steadily with increasing frequency. 
In contrast, optical phonons above $\sim$8~THz exhibit significantly shorter lifetimes, on the order of 5–30~ps, due to the increased scattering phase space. 
These trends are robust across all the employed models.

The 4$^{\mathrm{ph}}$ lifetimes are notably longer, with acoustic modes reaching up to 7500~ps and optical branches generally exceeding 1000~ps. 
This disparity indicates that 4$^{\mathrm{ph}}$ scattering processes, despite their higher-order nature, act on longer timescales and contribute less to resistive scattering in the low-to-intermediate frequency range. 
MoSe$_2$ displays similar qualitative behavior (Figures~S4 and S5), though the lifetimes are uniformly shorter. 
In the 3$^{\mathrm{ph}}$ case, lifetimes do not exceed 200~ps, reflecting the reduced phonon group velocities due to the heavier selenium atoms. 
For 4$^{\mathrm{ph}}$scattering in MoSe$_2$, lifetimes are also suppressed, typically remaining below 5000~ps, with especially short values ($<$1500~ps) for mid-to-high frequency optical modes. 
This suggests that quartic anharmonicity contributes more significantly to phonon scattering in MoSe$_2$ compared to MoS$_2$.

Overall, our results show that while 4$^{\mathrm{ph}}$ processes are characterized by much longer lifetimes than their 3$^{\mathrm{ph}}$ counterparts, their cumulative effect (particularly at elevated temperatures and for high-frequency optical modes) can still be substantial. 
The relative trends between MoS$_2$ and MoSe$_2$ align well with their experimentally and theoretically known thermal conductivities, highlighting the importance of including four-phonon interactions for an accurate and comprehensive description of intrinsic thermal transport. 
Moreover, the consistent behavior observed across different MLFF frameworks underscores their capability to reliably capture both cubic and quartic anharmonic effects, provided that the models are sufficiently trained and converged.

\begin{figure*}[h!]
    \centering
    \includegraphics[width=0.9\textwidth]{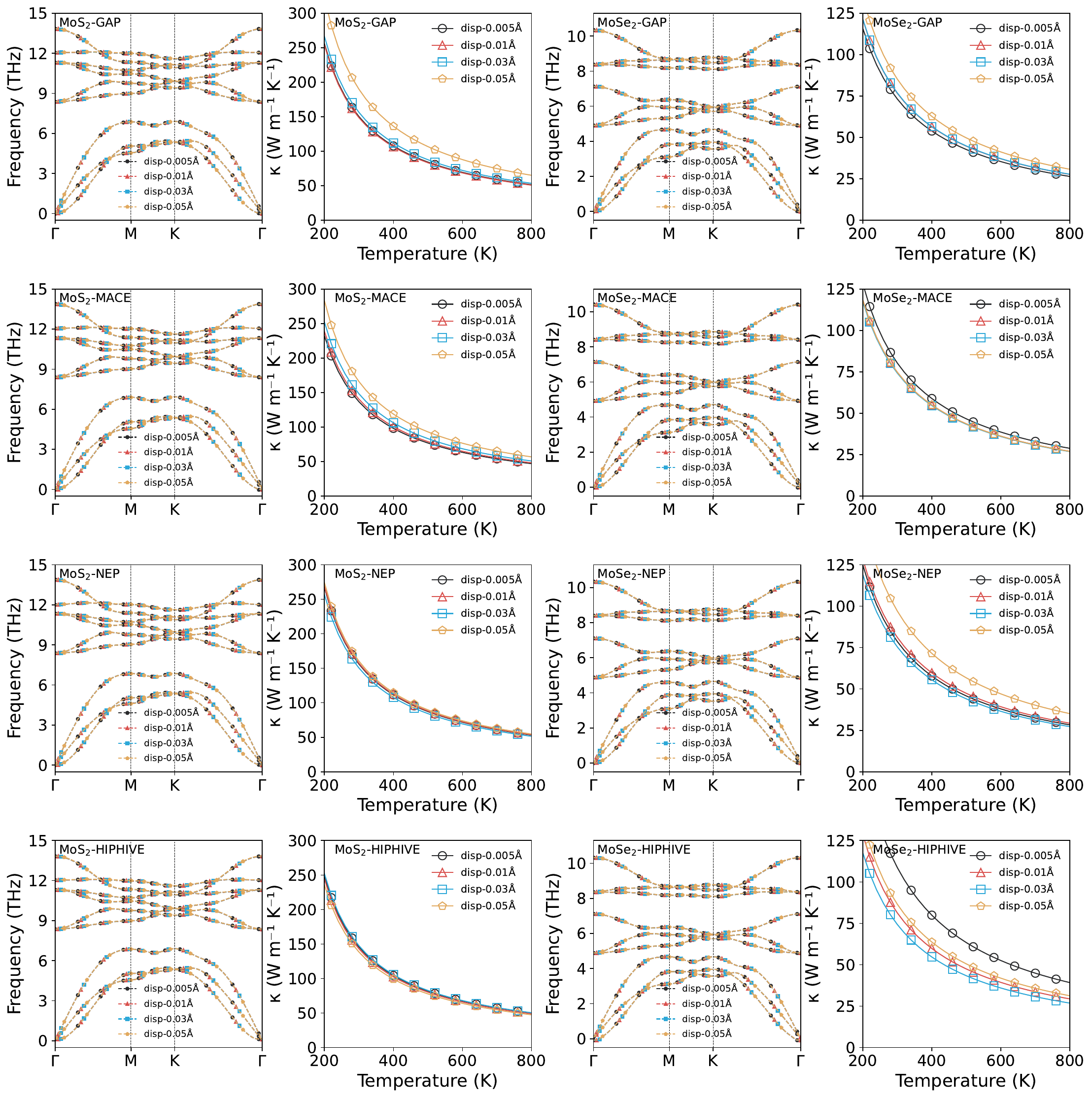}
    \caption{Phonon dispersion relations and lattice thermal conductivity as a function of temperature for monolayer MoS$_2$ and MoSe$_2$, obtained using displacement amplitudes of 0.005, 0.01, 0.03, and 0.05 Å. Force constants are constructed up to the 13$^{\text{th}}$ nearest neighbors (NN).}
    \label{fig:diff_disp}
\end{figure*}

As previously discussed, we employed an atomic displacement of 0.03 {\AA} to compute the force constants used in both phonon frequency and phonon-phonon scattering rate calculations. 
While DFT-based forces are generally robust across a range of displacement magnitudes (typically between 0.005 and 0.04 Å) and yield consistent results, the accuracy of MLFFs in the finite displacement method (FDM) can be more sensitive to the chosen displacement. 
In particular, MLFFs tend to produce significant force prediction errors at very small displacement values.

To systematically assess this sensitivity, we performed third-order interatomic force constant (IFC) calculations for the considered monolayers using a range of displacement magnitudes (0.005, 0.01, 0.03, and 0.05~\AA).
In each case, the same displacement magnitude was employed for both second- and third-order IFC calculations to ensure consistency. 
As shown in Figure~\ref{fig:diff_disp}, the phonon dispersion relations remain essentially unchanged across the tested displacements, exhibiting excellent agreement throughout the Brillouin zone for both materials. 
In contrast, the calculated $\kappa$ values display a more pronounced dependence on displacement magnitude. 
For the smallest (0.005~\AA) and largest (0.05~\AA) displacements, the $\kappa$–$T$ curves deviate noticeably, with the magnitude and direction of the variation depending on both the material and the employed MLFFs. 
These differences reflect the interplay between how each potential captures interatomic forces under very small or relatively large atomic perturbations in the finite-displacement framework, as well as intrinsic differences in the lattice dynamics of MoS$_2$ and MoSe$_2$—arising from their distinct atomic masses, bond strengths, and phonon spectra. 
By contrast, intermediate displacements of 0.01–0.03~\AA\ yield nearly identical $\kappa$–$T$ profiles for all tested models, indicating an optimal balance between numerical accuracy and the avoidance of artifacts associated with extreme displacement values. 
This optimal range is consistently observed for GAP, MACE, NEP, and \textsc{HIPHIVE}, suggesting that it is largely independent of the types of MLFF.

Compared to DFT reference thermal conductivities at 300 K (153 Wm$^{-1}$K$^{-1}$ for MoS$_2$ and 74 Wm$^{-1}$K$^{-1}$ for MoSe$_2$; see Table~\ref{tab:tc_table_compact_mlff}), displacements in the range of 0.01–0.03 {\AA} yield $\kappa$ values within 10\% deviation for most MLFF models. 
This indicates that the 0.01–0.03 {\AA} range offers an optimal trade-off between numerical stability and physical reliability. 
In contrast, displacements of 0.05 {\AA} result in significantly larger deviations—exceeding 25\% in some cases—notably for the MoS$_2$-GAP and MoSe$_2$-NEP models. 
These findings highlight the critical importance of displacement magnitude selection in FDM-based workflows, particularly when evaluating third-order IFCs for phonon-phonon scattering and thermal transport calculations.

\begin{figure*}[h!]
    \centering
    \includegraphics[width=14cm]{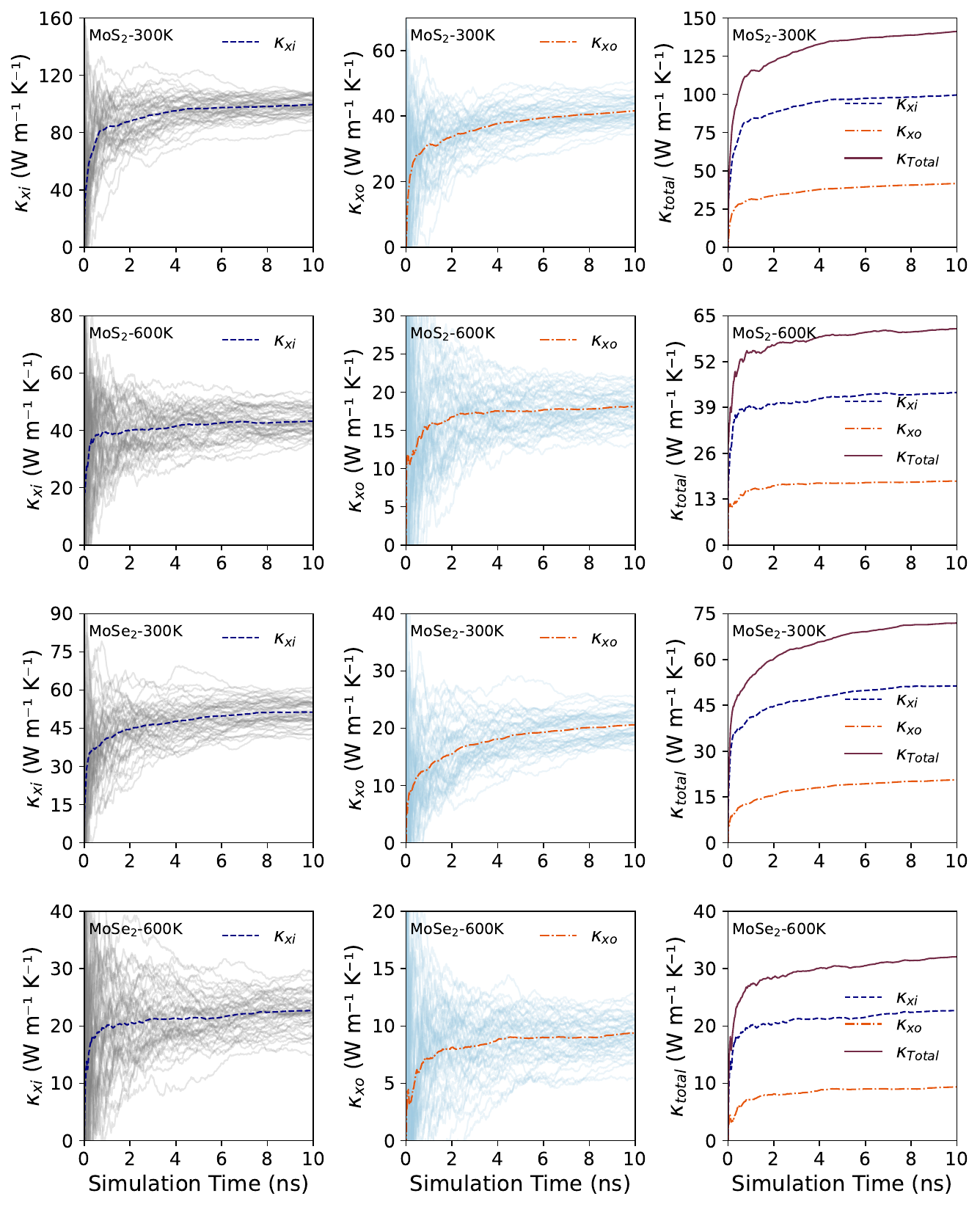}
    \caption{Lattice thermal conductivity components ($\kappa_{xi}$, $\kappa_{xo}$, $\kappa_{\mathrm{total}}$) of monolayer MoS$_2$ and MoSe$_2$ at 300~K and 600~K, obtained from HNEMD simulations. Here, $\kappa_{xi}$ and $\kappa_{xo}$ represent the in-plane and out-of-plane phonon contributions to the thermal conductivity along the $x$-axis, respectively, while $\kappa_{\mathrm{total}}$ is their sum. All values are plotted as a function of simulation time.}
    \label{fig:hnemd}
\end{figure*}

\subsection{Thermal transport properties: HNEMD simulations}
To further evaluate the reliability of the lattice thermal conductivity values obtained from PBTE solutions using DFT and MLFFs, we carried out homogeneous nonequilibrium molecular dynamics (HNEMD) simulations (see Ref.~\cite{Fan2019} for further details). 
This heat current based dynamical evaluation yields $\kappa_{\mathrm{total}}$ values of 141 and 61~$\text{Wm}^{-1}\text{K}^{-1}$ for MoS$_2$ at 300 and 600~K, respectively, and 72 and 32~$\text{Wm}^{-1}\text{K}^{-1}$ for MoSe$_2$ at the same temperatures, as depicted in Figure~\ref{fig:hnemd}. 
The in-plane, $\kappa_{xi}$ and out-of-plane, $\kappa_{xo}$ phonon contribution to the total conductivity along the $x$-direction, $\kappa_{Total}$ are calculated separately within this approach. The time evolution profiles show that $\kappa_{xi}$ dominates the total thermal conductivity, while $\kappa_{xo}$ remains notably smaller in all cases. This behavior is physically expected, because in-plane (LA/TA) phonons have higher group velocities due to the stiffness of covalent bonds in the plane and their lifetimes are comparatively longer. 
In contrast, out-of-plane (ZA) phonons are softer modes with quadratic dispersion near the $\Gamma$ point, resulting in lower group velocities and stronger anharmonic scattering, which together limit their contribution to heat transport.

These MD results serve as independent benchmarks for validating MLFF-based predictions. 
In particular, the fourth-order–corrected $\kappa$ values from GAP (140~$\text{Wm}^{-1}\text{K}^{-1}$), MACE (138~$\text{Wm}^{-1}\text{K}^{-1}$), and NEP (137~$\text{Wm}^{-1}\text{K}^{-1}$) are in excellent agreement with the HNEMD value of 142~$\text{Wm}^{-1}\text{K}^{-1}$ for MoS$_2$ at 300~K, deviating by less than 3\%. 
A similarly close match is observed for MoSe$_2$, where GAP (72~$\text{Wm}^{-1}\text{K}^{-1}$), MACE (70~$\text{Wm}^{-1}\text{K}^{-1}$), and NEP (70~$\text{Wm}^{-1}\text{K}^{-1}$) all reproduce the HNEMD value of 72~$\text{Wm}^{-1}\text{K}^{-1}$ within the margin of statistical uncertainty.
This high level of consistency reinforces the robustness of our MLFFs-based PBTE framework and further suggests that the dramatic $\kappa$ reductions reported in some earlier monolayer MoS$_2$ studies after including fourth-order scattering are likely overestimated.

\begin{figure*}[ht!]
    \centering
\includegraphics[width=0.9\textwidth]{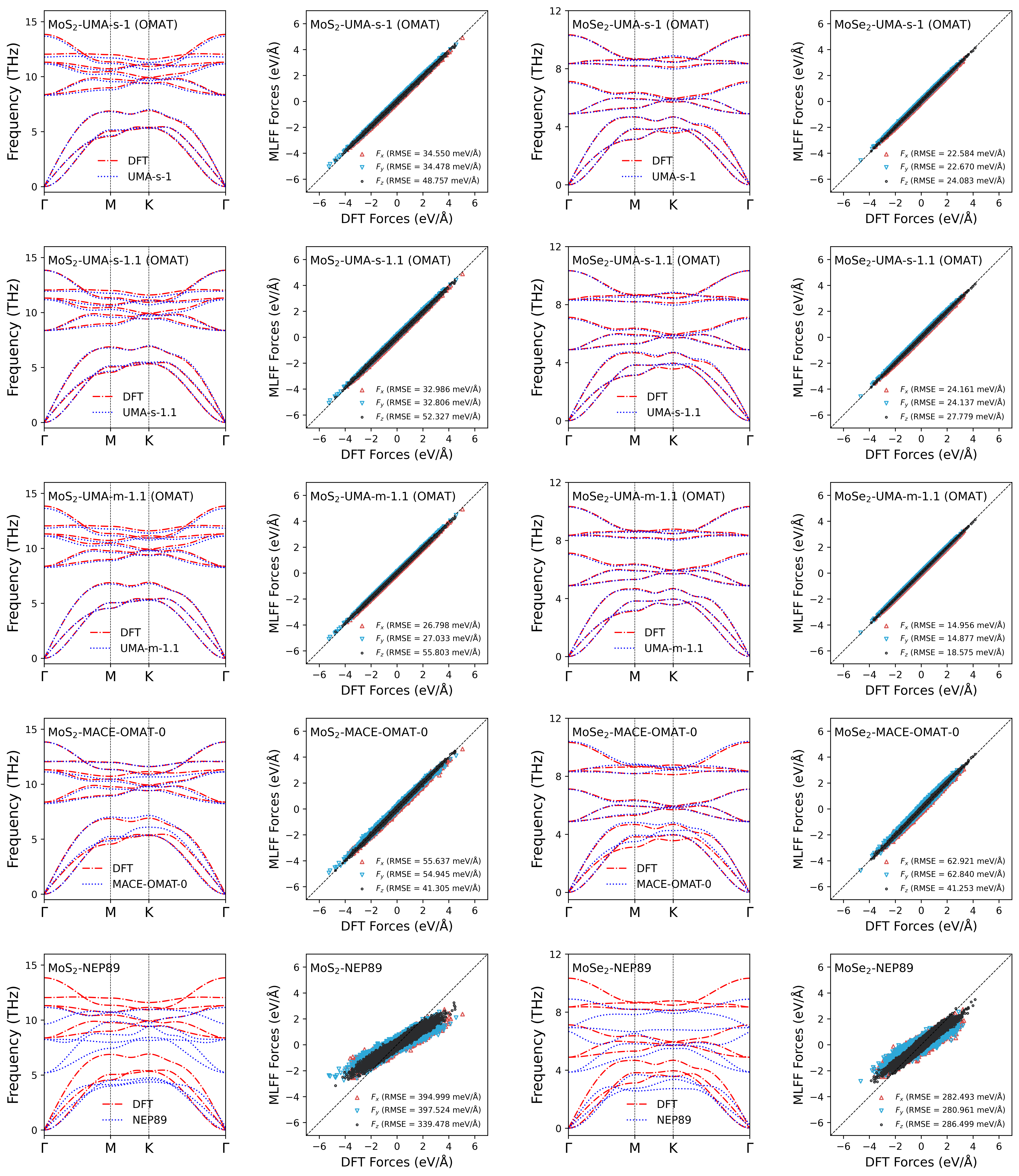}
    \caption{Phonon dispersion relations and atomic force predictions for monolayer MoS$_2$ and MoSe$_2$, evaluated using various universal MLFFs: UMA-s-1, UMA-s-1.1, UMA-m-1.1, MACE-OMAT-0, and NEP89. For each model, phonon spectra are compared with density functional theory (DFT) results, and the accuracy of predicted atomic forces is quantified by the root-mean-square error (RMSE) with respect to DFT values using independent test datasets.}
    \label{fig:univ_pots}
\end{figure*}

\subsection{Universal potentials}

The development of \emph{universal interatomic potentials}, also referred to as pretrained foundation models, has recently emerged as a promising route toward transferable and data-efficient models for atomistic simulations across diverse material classes.
Unlike traditional machine-learned force fields (MLFFs), which are trained for specific systems, these models are pretrained on large chemically diverse datasets and can be adapted via fine-tuning. 
Representative examples include MACE-OMAT-0~\cite{batatia2023foundation}, UMA~\cite{wood2025umafamilyuniversalmodels} developed by Meta AI, and NEP89~\cite{liang2025nep89universalneuroevolutionpotential}. 
Their goal is to generalize atomic interactions and enable reliable predictions of energies, forces, and virials even in previously unseen systems.

We evaluated the transferability of these potentials in their released form, without any fine-tuning, by testing their ability to generate accurate interatomic force constants (FCs) for monolayer MoS$_2$ and MoSe$_2$. The harmonic properties were first examined through phonon dispersions derived from second-order FCs and force–parity plots benchmarked against DFT (see Figure~\ref{fig:univ_pots}). 
UMA-s-1.1 and UMA-m-1.1 provided the closest agreement across the Brillouin zone, while MACE-OMAT-0 showed a modest high-frequency bias.
NEP89 produced large force errors and distorted dispersions and was therefore excluded from transport calculations.

The anharmonic response was assessed by computing the lattice thermal conductivity at 300 and 600~K using Peierls-Boltzmann transport equation solutions with third- and fourth-order FCs. 
Consistent with the trained MLFFs, the third-order force constants were averaged over neighbor cutoffs ranging from the 13$^{\text{th}}$ to the 18$^{\text{th}}$ shell, whereas the fourth-order force constants were calculated using a displacement amplitude of 0.04~\AA{} and averaged over the 6$^{\text{th}}$–10$^{\text{th}}$ shells. 
The corresponding results are summarized in Table~\ref{tab:tc_table_compact_mlff_univ_pots}.

\begin{table}[htbp]
\centering
\caption{Lattice thermal conductivity values (Wm$^{-1}$K$^{-1}$) of monolayer MoS$_2$ and MoSe$_2$ at 300 and 600 K, calculated using various universal potentials. Results labeled 3$^{\mathrm{ph}}$ are obtained from PBTE including only three-phonon scattering, with $\kappa$ values averaged over the 13$^{\text{th}}$–18$^{\text{th}}$ nearest-neighbor (NN) range to ensure convergence with DFT. The 4$^{\mathrm{ph}}$ results include both three- and four-phonon processes, where three-phonon contributions are obtained for the 13$^{\text{th}}$ NN and four-phonon contributions are averaged over the 6$^{\text{th}}$, 8$^{\text{th}}$, and 10$^{\text{th}}$ NN ranges.}
\label{tab:tc_table_compact_mlff_univ_pots}

\scriptsize
\setlength{\tabcolsep}{5pt}
\begin{tabular}{|l|l|r|r|r|r|}
\hline
\multirow{2}{*}{Material} & \multirow{2}{*}{Method}
& \multicolumn{2}{c|}{300~K} & \multicolumn{2}{c|}{600~K} \\
\cline{3-6}
& & 3$^{\mathrm{ph}}$& 4$^{\mathrm{ph}}$& 3$^{\mathrm{ph}}$& 4$^{\mathrm{ph}}$\\
\hline
\multirow{6}{*}{MoS\textsubscript{2}}
& DFT (ref)        & 153 (ref)       & \textemdash      & 71 (ref)       & \textemdash \\
& HNEMD (ref)      & \textemdash      & 141 (ref)        & \textemdash     & 61 (ref) \\
& UMA-s-1          & 143 [$-6.54\%$]  & 114 [$-19.15\%$] & 66 [$-7.04\%$]  & 47 [$-22.95\%$] \\
& UMA-s-1.1        & 166 [$+8.50\%$]  & 140 [$-0.71\%$]  & 77 [$+8.45\%$]  & 56 [$-8.20\%$] \\
& UMA-m-1.1        & 151 [$-1.31\%$]  & 131 [$-7.09\%$]  & 70 [$-1.41\%$]  & 54 [$-11.48\%$] \\
& MACE-OMAT-0      & 174 [$+13.73\%$] & 158 [$+12.06\%$] & 80 [$+12.68\%$] & 67 [$+9.84\%$] \\
\hline
\multirow{6}{*}{MoSe\textsubscript{2}}
& DFT (ref)        & 74 (ref)         & \textemdash      & 36 (ref)       & \textemdash \\
& HNEMD (ref)      & \textemdash       & 72 (ref)         & \textemdash     & 32 (ref) \\
& UMA-s-1          & 75 [$+1.35\%$]   & 61 [$-15.28\%$]  & 39 [$+8.33\%$]  & 25 [$-21.88\%$] \\
& UMA-s-1.1        & 103 [$+39.19\%$] & 73 [$+1.39\%$]   & 49 [$+36.11\%$] & 30 [$-6.25\%$] \\
& UMA-m-1.1        & 80 [$+8.11\%$]   & 66 [$-8.33\%$]   & 39 [$+8.33\%$]  & 28 [$-12.50\%$] \\
& MACE-OMAT-0      & 70 [$-5.41\%$]   & 62 [$-13.89\%$]  & 34 [$-5.56\%$]  & 27 [$-15.62\%$] \\
\hline
\end{tabular}
\normalsize
\end{table}

A closer inspection of the relative errors highlights clear trends in all evaluated models. 
For MoS$_2$, UMA-s-1.1 reproduces the three-phonon DFT value within $+8.5\%$, while its four-phonon 
prediction differs from the HNEMD benchmark by less than $1\%$, indicating excellent agreement. 
UMA-m-1.1 also performs consistently well, with deviations within $-1.3\%$ (3$^{\mathrm{ph}}$) and $-7.1\%$ (4$^{\mathrm{ph}}$). 
By contrast, UMA-s-1 systematically underestimates $\kappa$, reaching errors up to $-23\%$ for the 
four-phonon case at 600~K, while MACE-OMAT-0 tends to overestimate, exceeding $+12\%$ in several cases. 
For MoSe$_2$, the variability is more pronounced: UMA-s-1.1 overpredicts three-phonon conductivities by nearly $40\%$, although its four-phonon values remain within $+1.4\%$ of HNEMD. UMA-m-1.1 maintains 
errors below $10\%$, whereas MACE-OMAT-0 underestimates four-phonon results by $-14$ to $-16\%$. 
These patterns demonstrate that while certain universal potentials can reach near-DFT accuracy, model-to-model fluctuations on the order of $10$–$20\%$ remain common.

Overall, UMA-s-1.1 and UMA-m-1.1 delivered harmonic and anharmonic transport properties competitive with task-specific MLFFs even without fine-tuning, but significant variability across models was observed. 
UMA-s-1 underestimated conductivities, MACE-OMAT-0 showed pronounced deviations, and NEP89 was excluded 
due to large force errors. 
Notably, all models except NEP89 reproduced the quadratic acoustic dispersion near the $\Gamma$ point, capturing long-wavelength phonons and hence dynamical stability. 
Since acoustic modes are the main heat carriers in semiconductors, this accuracy is critical for reliable $\kappa$ predictions. 
While universal models are not yet a substitute for custom MLFFs in high-precision studies, their performance in completely unseen systems highlights their potential as efficient starting points for fine-tuned applications.

\section{Computational Times}

Since high-order anharmonic calculations are computationally demanding, the practical value of MLFFs and universal potentials depends not only on accuracy but also on computational efficiency. Table~\ref{tab:comp_times} reports the average time required to compute atomic forces for the considered methods.

\begin{table}[htbp]
\centering
\caption{Average computation times for evaluating atomic forces using different methods. MLFF timings are averaged over 100 structures.}
\label{tab:comp_times}
\scriptsize
\begin{tabular}{|l|c|c|}
\hline
Method & Comp. Time & Hardware \\
\hline
GAP        & 1.45 s  & CPU via ASE \\
MACE-GPU   & 0.37 s  & 1 $\times$ A100 GPU \\
MACE-CPU   & 4.96 s  & CPU via ASE \\
NEP        & 0.05 s  & CPU via ASE \\
\textsc{HIPHIVE}    & 0.93 s  & CPU via ASE \\
DFT        & 90 min  & 4 $\times$ A100 GPUs \\
\hline
\end{tabular}
\normalsize
\end{table}

All MLFFs achieve speedups of several orders of magnitude compared to DFT, reducing the force evaluation time from hours to milliseconds. 
Among them, GPU-accelerated MACE delivers the highest performance overall, while NEP offers the lowest latency on CPUs, making it particularly attractive for large-scale MD. 
GAP and \textsc{HIPHIVE} provide balanced accuracy–efficiency trade-offs with stable throughput, and CPU-based MACE remains competitive despite being slower. 
By contrast, DFT is prohibitively expensive, requiring $\sim90$ minutes per structure even on multiple GPUs.

These efficiency gains are not merely technical: they make it feasible to include extended neighbor cutoffs and fourth-order IFCs in PBTE calculations, and they also enable long-time molecular dynamics simulations that are beyond the reach of first-principles methods. 
In practice, the ability to choose between GPU-accelerated and CPU-based MLFFs provides flexibility depending on the available resources and the scale of the targeted problem, ranging from high-throughput screening to detailed transport modeling.

\section{Conclusions}
The characterization of thermal transport in TMDs, and low-dimensional materials more broadly, has been persistently hampered by large discrepancies in/between calculated and measured lattice thermal conductivities. In this work, we combined state-of-the-art MLFFs with high-accuracy ab initio calculations to resolve these long-standing inconsistencies. Our systematic evaluation, which integrates MLFFs with both the PBTE and molecular dynamics simulations, demonstrates that this approach provides a reliable and computationally efficient reference for reconciling conflicting findings in the literature. All the tested MLFFs (GAP, MACE, NEP, and \textsc{HIPHIVE}) successfully capture the essential physics of lattice thermal transport. Notably, the NEP model achieves good fidelity to first-principles benchmarks while simultaneously delivering an order-of-magnitude improvement in simulation speed when compared with the other tested alternatives. This powerful combination of accuracy and efficiency marks a significant step forward, enabling predictive and high-throughput studies of thermal transport that were previously intractable.

Beyond benchmarking, our analysis clarifies the role of higher-order phonon–phonon interactions. 
We found that quartic anharmonicity introduces only moderate corrections, in contrast to earlier predictions of drastic reductions, yielding intrinsic conductivity limits of MoS$_2$ and MoSe$_2$ at approximately 140 and 70 Wm$^{-1}$K$^{-1}$, respectively. 
This refinement highlights the need for convergence-tested methodologies and illustrates the value of MLFFs in extending the accuracy of first-principles calculations to regimes that were previously inaccessible.

Looking ahead, MLFFs emerge not only as efficient substitutes for density functional theory in quantitative transport studies but also as enabling tools for broader applications. 
The encouraging performance of transferable universal potentials points toward a pathway for general-purpose force fields that can accelerate high-throughput exploration while retaining physical fidelity. 
Such advances set the stage for the solution of more complex thermal transport problems in two-dimensional systems, including the critical influence of defects, interfaces, and nanoscale heterogeneity.

\newpage

\section*{Acknowledgments}
This work was partially supported by the Research Foundation–Flanders (FWO–Vl). The computational resources and services for this work were provided by the High Performance and Grid Computing Center (TRGrid Infrastructure) of TUBITAK ULAKBIM and the National Center for High Performance Computing (UHeM) of Istanbul Technical University, and VSC (Flemish Supercomputer Center), funded by the FWO and the Flemish Government – department EWI. 
This research also used resources of the Argonne Leadership Computing Facility, a U.S. Department of Energy (DOE) Office of Science user facility at Argonne National Laboratory and is based on research supported by the U.S. DOE Office of Science-Advanced Scientific Computing Research Program, under Contract No. DE-AC02-06CH11357.

\section*{data availability statement}
The MLFFs and the training test and validation data that support the findings of this study are openly available in Zenodo at http://doi.org/, reference number [reference number will be provided after acceptance]. Also, any additional data that support the findings of this study are available on request from the corresponding author.

\newpage

\bibliographystyle{unsrt}
\bibliography{ref}

\end{document}


\title{\textbf{Supplementary Materials\\ for\\ Revisiting the Thermal Conductivity Limits of MoS$_2$ and MoSe$_2$: Insights from High-Order Anharmonic Lattice Dynamics with Machine Learning Potentials}}

\author{Tu\u{g}bey Kocaba\c{s}}
\affiliation{Department of Advanced Technologies, Eskisehir Technical University, 26555 Eskisehir, T\"{u}rkiye}
\author{Murat Ke\c{c}eli}
\affiliation{Computational Science Division, Argonne National Laboratory, Lemont IL 60517, USA}
\author{Tanju Gürel}
\affiliation{Department of Physics, Tekirdağ Namik Kemal University, Tekirdağ TR 59030, T\"{u}rkiye}
\author{Milorad V. Milo\v{s}evi\'c}
\affiliation{Department of Physics \& NANOlab Center of Excellence, University of Antwerp, Groenenborgerlaan 171, B-2020 Antwerp, Belgium}
\author{Cem Sevik}
\affiliation{Department of Physics \& NANOlab Center of Excellence, University of Antwerp, Groenenborgerlaan 171, B-2020 Antwerp, Belgium}

\date{\today}

\maketitle

\newpage

\begin{table}[htbp]
\centering
\caption{Fitting hyperparameters of the GAP model for MoS$_2$ and MoSe$_2$.}
\label{tab:gap_hyperparams}
\scriptsize
\setlength{\tabcolsep}{4pt}
\renewcommand{\arraystretch}{1.3}
\begin{tabular}{|l|l|c|c|c|p{6cm}|}
\hline
Block & Descriptor & Cutoff (\AA) & $\delta$ & $n_{\text{sparse}}$ & Notes \\
\hline
\multicolumn{6}{|l|}{\textbf{General:} $sparse\_jitter=10^{-12}$, 
$e_0$(Mo:$-4.5876$, S:$-0.9546$, Se:$-0.7836$) eV, 
default $\sigma=\{10^{-8},10^{-8},10^{-4},0\}$} \\
\hline
2B & distance\_2b & 6.0 & 1.0 & 50 & ARD\_SE kernel, $\theta=1.0$; transition width $0.5$~\AA \\
\hline
3B & angle\_3b (all triplets) & 4.5 & 0.05 & 100 & ARD\_SE kernel, $\theta=1.0$; transition width $0.5$~\AA \\
\hline
SOAP-turbo & Mo- and S/Se-centered & 7.0/7.5 & 0.1 & 10000 & $l_{\max}=8$, $\alpha_{\max}=\{8,8\}$, $\sigma_r=\sigma_t=0.7$, $\zeta=10$, basis=poly3gauss \\
\hline
\end{tabular}
\normalsize
\end{table}

\begin{table}[htbp]
\centering
\caption{Training hyperparameters of the MACE model for MoS$_2$ and MoSe$_2$.}
\label{tab:mace_hyperparams}
\scriptsize
\setlength{\tabcolsep}{4pt}
\renewcommand{\arraystretch}{1.2}
\begin{tabular}{|l|l|}
\hline
Parameter & Value \\
\hline
Model type & MACE \\
Radial cutoff $r_{\max}$ (\AA) & 8.0 \\
Max angular momentum $L_{\max}$ & 3 \\
Correlation order & 4 \\
Hidden irreps & 128$\times$0e + 128$\times$1o + 128$\times$2e \\
Number of interactions & 3 \\
Optimizer options & AMSGrad, learning rate 0.005 \\
Max epochs & 1000 \\
EMA decay & 0.9999 \\
SWA (start / lr) & 750 / 0.001 \\
Weights (Energy / Forces / Stress) & 1.0 / 200.0 / 1.0 \\
Reference energies ($E_0$) & Mo: $-4.5876$, S: $-0.9546$, Se:$-0.7836$ eV \\
Error metric & Per-atom RMSE \\
\hline
\end{tabular}
\normalsize
\end{table}

\begin{table}[htbp]
\centering
\caption{Training hyperparameters of the NEP model for MoS$_2$ and MoSe$_2$}
\label{tab:nep_hyperparams}
\scriptsize
\setlength{\tabcolsep}{5pt}
\renewcommand{\arraystretch}{1.2}
\begin{tabular}{|l|l|}
\hline
Parameter & Value \\
\hline
Version & NEP-4 \\
Cutoffs & Radial: 10 \AA, Angular: 8 \AA \\
Typewise factors & Radial: 2.5, Angular: 2 \\
Radial / Angular basis size & 8 / 8 \\
$n_{\max}^{\mathrm{radial}}$ / $n_{\max}^{\mathrm{angular}}$ & 6 / 6 \\
$l_{\max}$ (3-,4-,5-body) & 6 / 2 / 1 \\
Neurons per layer & 100 \\
Batch size & 3000 \\
Max generations & 100000 \\
Population size & 50 \\
Loss weights & $\lambda_e=1$, $\lambda_f=300$, $\lambda_v=1$, $\lambda_{\mathrm{shear}}=5$ \\
Other & $\lambda_1=\lambda_2=0.0822$, force\_delta=0.001 \\
\hline
\end{tabular}
\normalsize
\end{table}

\begin{figure}[h!]
    \centering
    \includegraphics[width=0.6\columnwidth]{figures/phonons.pdf}
    \caption{Phonon dispersion relations of monolayer MoS$_2$ (up) and MoSe$_2$ (down), computed using DFT and MLFFs: GAP, MACE, NEP, and \textsc{HIPHIVE}. The comparison highlights the accuracy of MLFFs in reproducing the phonon spectra obtained from DFT.}
    \label{fig:phonons}
\end{figure}

\begin{figure}[h!]
    \centering
    \includegraphics[width=1\columnwidth]{supp_figures/lifetime_3ph_MoS2.pdf}
    \caption{Three-phonon ($3^{\mathrm{ph}}$) lifetimes of all phonon modes in monolayer MoS$_2$ at $T = 300$~K, plotted as a function of phonon frequency, as obtained from DFT, GAP, MACE, NEP, and HIPHIVE.}
    \label{fig:mos2_lifetime_3ph}
\end{figure}

\begin{figure}[h!]
    \centering
    \includegraphics[width=1\columnwidth]{supp_figures/lifetime_3ph_MoSe2.pdf}
    \caption{Three-phonon ($3^{\mathrm{ph}}$) lifetimes of all phonon modes in monolayer MoSe$_2$ at $T = 300$~K, plotted as a function of phonon frequency, as obtained from DFT, GAP, MACE, NEP, and HIPHIVE.}
    \label{fig:mose2_lifetime_3ph}
\end{figure}

\begin{figure}[h!]
    \centering
    \includegraphics[width=1\columnwidth]{supp_figures/lifetime_4ph_MoSe2.pdf}
    \caption{Four-phonon ($4^{\mathrm{ph}}$) lifetimes of all phonon modes in monolayer MoSe$_2$ at $T = 300$~K, plotted as a function of phonon frequency, as obtained from GAP, MACE, NEP, and HIPHIVE.}
    \label{fig:mose2_lifetime_4ph}
\end{figure}

\begin{figure}[h!]
    \centering
    \includegraphics[width=1\columnwidth]{supp_figures/lifetime_4ph_MoS2.pdf}
    \caption{Four-phonon ($4^{\mathrm{ph}}$) lifetimes of all phonon modes in monolayer MoS$_2$ at $T = 300$~K, plotted as a function of phonon frequency, as obtained from GAP, MACE, NEP, and HIPHIVE.}
    \label{fig:mos2_lifetime_4ph}
\end{figure}

\begin{figure}[h!]
    \centering
    \includegraphics[width=0.7\columnwidth]{supp_figures/nn_vs_disp_4th_mlff_MoS2.pdf}
    \caption{Convergence of the lattice thermal conductivity ($\kappa$) of monolayer MoS$_2$ with respect to the nearest-neighbor (nn) cutoff in the \emph{fourth-order} force-constant calculations, for different displacement amplitudes (0.01–0.05~\AA). Results are obtained from solutions of the phonon Boltzmann transport equation (BTE) using machine-learned force fields (GAP, MACE, NEP, and HiPhive), with the \emph{third-order} force constants fixed using a 13$^{\mathrm{th}}$-nearest-neighbor cutoff.}
    \label{fig:mos2_4ph_nn}
\end{figure}

\begin{figure}[h!]
    \centering
    \includegraphics[width=0.7\columnwidth]{supp_figures/nn_vs_disp_4th_mlff_MoSe2.pdf}
    \caption{Convergence of the lattice thermal conductivity ($\kappa$) of monolayer MoSe$_2$ with respect to the nearest-neighbor (nn) cutoff in the \emph{fourth-order} force-constant calculations, for different displacement amplitudes (0.01–0.05~\AA). Results are obtained from solutions of the phonon Boltzmann transport equation (BTE) using machine-learned force fields (GAP, MACE, NEP, and HiPhive), with the \emph{third-order} force constants fixed using a 13$^{\mathrm{th}}$-nearest-neighbor cutoff.}
    \label{fig:mose2_4ph_nn}
\end{figure}

\begin{figure}[h!]
    \centering
    \includegraphics[width=0.7\columnwidth]{supp_figures/nn_tc_mlff_4th.pdf}
    \caption{Lattice thermal conductivity ($\kappa$) of monolayer MoS$_2$ and MoSe$_2$ at room temperature, obtained from solutions of the phonon Boltzmann transport equation (BTE), as a function of the nearest-neighbor (nn) cutoff used in the \emph{fourth-order} force-constant calculations. Results are shown for machine-learned force fields (GAP, MACE, NEP, and HiPhive), where the \emph{third-order} force constants were fixed using a 13$^{\mathrm{th}}$-nearest-neighbor cutoff. The \emph{fourth-order} force constants were computed using a displacement amplitude of 0.04~\AA.}
    \label{fig:mos2_4ph_nn}
\end{figure}

\begin{figure}[h!]
    \centering
    \includegraphics[width=0.7\columnwidth]{supp_figures/nn_tc_univ_pots_3rd.pdf}
    \caption{Lattice thermal conductivity ($\kappa$) of monolayer MoS$_2$ and MoSe$_2$ at room temperature as a function of the nearest-neighbor (nn) cutoff used in the third-order force-constant calculations. Curves compare DFT with universal pretrained potentials (UMA-s-1, UMA-s-1.1, UMA-m-1.1, and MACE-OMAT-0), considering only three-phonon (3$^{\mathrm{ph}}$) scattering.}
    \label{fig:univ_pots_nn_3rd}
\end{figure}

\bibliographystyle{unsrt}
\bibliography{ref}